# Strain-Driven Structure-Ferroelectricity Relationship in hexagonal $TbMnO_3$ Films

R. Mandal[1,2], M. Hirsbrunner[1], V. Roddatis[3], L. Schüler[1], U. Roß[3], S. Merten[1], and V. Moshnyaga[1*]

*[1]Erstes Physikalisches Institut, Georg-August-Universität Göttingen, Friedrich-Hund-Platz 1, 37077 Göttingen, Germany*

*[2]Department of Physics, Indian Institute of Science Education and Research, Pune 411008, India*

*[3]Institut für Materialphysik, Georg-August-Universität Göttingen, Friedrich-Hund-Platz 1, 37077 Göttingen, Germany*

**E-mail: vmosnea@gwdg.de*

**Keywords:** rare earth manganites, epitaxy, hexagonal $TbMnO_3$, transmission electron microscopy, ferroelectricity, Raman spectroscopy, optical ellipsometry.

Thin films and heterostructures of hexagonal manganites as promising multiferroic materials have attracted a considerable interest recently. We report structural transformations of high quality epitaxial h-TMO/YSZ(111) films, analyzed by means of various characterization techniques. A phase transition from $P6_3mc$ to $P6_3mcm$ structure at $T_C$~800 K was observed by temperature dependent Raman spectroscopy and optical ellipsometry. The latter probing directly electronic system, indicates its modification at the structural phase transition likely due to charge transfer from oxygen to Mn. In situ transmission electron microscopy (TEM) of the lamella samples displayed an irreversible $P6_3mc$-$P6_3mcm$ transformation and vanishing of ferroelectric domains already at 410 K. After the temperature cycling (300K-1300K-300K) the room temperature TEM of h-TMO films revealed an inhomogeneous microstructure, containing ferroelectric and paraelectric nanodomains with $P6_3mc$ and $P6_3mcm$ structure, respectively. We point out a strong influence of stress relaxation, induced by temperature and by constrained sample geometry onto the structure and ferroelectricity in strain-stabilized h-TMO thin films.

## Introduction

Rare earth manganites, $REMnO_3$ with RE=La to Lu, manifest a family of strongly correlated electronic oxides with rich emerging properties, like high-temperature ferroelectricity, magnetic ordering and magneto-electric coupling[1,2]. Two distinct crystal structures, i.e. orthorhombic (Pnma) and hexagonal ($P6_3cm$), were revealed for large (La, Pr, Nd, Sm, Gd, Tb, Dy) and small (Y, Ho, Er, Tm, Yb, Lu) RE-cations, respectively. Both structural polymorphs demonstrate coupled ferroelectric and magnetic properties, although the origin and the emergence of ferroelectric ordering is drastically different. In the Pnma structure, the improper ferroelectricity with very low Curie temperatures, $T_C$~15K (o-$HoMnO_3$)[3], has a magnetoelastic origin; the modulation of magnetic structure results in a typically small spontaneous polarization, $P_s$=0.5 μC/cm$^2$ ($GdMnO_3$)[4]. In contrast, the ferroelectricity in the hexagonal phase originates from the geometric distortion yielding a very high $T_C$~1000 K[5,6] and much larger polarization, e.g. $P_s$=5.5 μC/cm$^2$ in h-$YMnO_3$[7,8]. For most of the hexagonal manganites ferroelectricity arises from the $K_3$ tilting of the $MnO_5$ bipyramids. The coupling of this distortion to the polar $\Gamma_2^-$ mode along with a buckling of RE-layer[9] results in an off-centrosymmetric displacement of oxygen and rare earth atoms. These structural modifications, having the form of buckling of atomic chains[10], can be directly visualized by means of a high resolution electron microscopy.

Being in the focus of condensed matter community for decades, the high-temperature structure/ferroelectric transitions in h-manganites were studied by different techniques. Recently H. Sim et al[11] have analyzed the high temperature X-ray diffraction (XRD) in bulk h-$YMnO_3$ and h-$ErMnO_3$ samples and proved the transition from the low temperature $P6_3cm$ structure to the high temperature $P6_3/mmc$ structure to be one-step, primarily caused by the RE-$O_P$ displacement of the $K_3$ phonon mode. Moreover, a nonlinear optical second harmonic generation (SHG) as a contact-free measurement[5,12-14] can probe the spontaneous polarization as well as its temperature dependence, allowing a determination of the ferroelectric $T_C$. However, as noted in Ref. 5 to avoid the inhomogeneity effects due to differently polarized ferroelectric domains a preparation of single-domain state by poling and, hence, the availability of a bottom electrode is necessary even for SHG study of a thin film. In addition, Raman scattering was also shown to be an effective nondestructive optical probe of structural phase transition in hexagonal manganites. Raman study on $YMnO_3$ bulk samples revealed that the $A_1$ symmetry of the dominating phonon mode at ~680 cm$^{-1}$ in the ferroelectric $P6_3cm$ structure changes to the $A_{1g}$ symmetry in the paraelectric $P6_3/mmc$ phase[15]. However, the studies of high temperature phase transitions in hexagonal films are very limited up to now: one can mention the observation of ferroelectric $T_C$~1020 K by monitoring the temperature behavior of the $A_1$ mode in the Raman spectra of the multiferroic 200 nm thick h-$LuFeO_3$ thin film[16].

Another problem with hexagonal multifrroic manganites $RMnO_3$ (R= Y, Ho, Er, Tm, Yb, Lu) showing geometrical ferroelectricity is the coupling between the structural and ferroelectric phase transitions, questioned in different studies[17-19]. Ghosh et al[17] pointed out the existence of a large temperature gap between the high temperature structural ($P6_3cm \Rightarrow P6_3/mmc$) transition, measured by X-ray diffraction, and the ferroelectric transition, i.e. the appearance of the spontaneous polarization in h-$LuMnO_3$. This gap, being maximal ~540 K for the smallest Lu cation and, hence, the most stable h-$LuMnO_3$, decreases down to 300 K for the border member h-$YMnO_3$[18]. As was shown by first principle calculations[19] the reason for large temperature gap could be the freezing-in of the $K_3$ phonon at structural phase transition and the resulting suppression of the ferroelectric polarization: the amplitude of the K3 phonon needs to increase critically by further cooling to provide its coupling to the polar $\Gamma_2^-$ mode.

Bulk $TbMnO_3$ (TMO) with relatively large Tb cation crystallizes in an orthorhombic structure and possesses a very small spontaneous polarization, $P_S$(10 K)=0.08μC/cm$^2$, below the ferroelectric $T_C$=27 K[20]. The Tb ion is in the middle of the rare earth row and its cation radius exceeds that of Y-cation, which is the border member of the thermodynamically stable hexagonal manganites. Thus, for TMO the thermodynamic formation energies of the h- and o-phases could be of comparable values, opening a way to shift the phase equilibrium towards the h-phase by applying special processing conditions during thin film preparation. The reports on the preparation of h-$TbMnO_3$ films are scarce and the obtained results are controversial. Namely, Lee et al.[21,22] demonstrated a stabilization of the hexagonal phase in TMO films grown by means of pulsed laser deposition (PLD) technique on Pt(111)/$Al_2O_3$(0001) and yttrium-stabilized zirconia (YSZ(111)) substrates. The films were found to be ferroelectric (FE) with $T_C$~60 K and for $T>T_C$ a transition into an anti-ferroelectric (AFE) phase was observed. Kim et al.[23] reported ferroelectricity at room temperature in a PLD grown h-TMO film and demonstrated the switching of the ferroelectric polarization by means of a piezo-force microscopy; no AFE phase was observed in contradiction with the previous report. By first principle calculations Kim et al.[23] predicted the ferroelectricity in h-TMO to have the same mechanism as in other hexagonal manganites. However, the information on ferroelectric domain structure as well as on the Curie temperature for this strain-stabilized h-phase is still absent and the microscopic origin of this FE phase was not clarified. A special interest to the h-TMO, compared to other h-manganites (R=Y-Lu), is motivated by its fewer thermodynamic stability, which allows one to expect a pronounced strain-driven ferroelectric and structural properties. Moreover, h-TMO was theoretically predicted to be an efficient photovoltaic material, in which an optimal bandgap and strong light absorption should result to the estimated 33 % of photovoltaic efficiency[24].

Here we report strain-stabilized h-TMO/YSZ(111) films grown by means of a metalorganic aerosol deposition (MAD) technique[25]. The films reveal a coherent epitaxial growth and possess perfect crystallinity as well as atomically sharp and flat film/substrate interfaces. The atomically resolved HAADF-STEM evidences the characteristic periodic buckling of Tb ions, indicating the presence of the FE domains with typical size of 5-20 nm, which are vertically aligned with respect to the film plane. The FE/PE transition at $T_C$~800 K was determined by means of temperature dependent Raman spectroscopy and optical ellipsometry.

## Experiments

The h-TMO films with a thickness d=50 nm have been grown by MAD technique[25] on commercial YSZ(111) substrates (Crystal GmbH) by using Tb- and Mn-acetylacetonates as precursors. Precursor solutions in dimethylformamide with concentration 0.02 M for Mn-precursor and molar ratios Tb/Mn=1.15-1.2 were used. The films were grown with the rate, v=6 nm/min, by spraying the precursor solution by dried compressed air onto the YSZ substrate heated to $T_{sub}$~900-1000°C in ambient atmosphere. After preparation the films were quenched down in 1 minute to 600°C and then cooled down to room temperature in 5 min. The substrate temperature was controlled by a pyrometer and by in situ optical ellipsometry.

XRD (Cu-$K_\alpha$ radiation, Θ-2Θ Bragg-Brentano geometry), rocking curves (ω-scans), asymmetric ϕ-scan and small-angle X-ray reflectivity (XRR) were used to characterize the structure and thickness of the films. Polarization dependent Raman spectroscopy was carried out at room temperature by using a confocal Raman microscope (LabRAM HR Evolution, Horiba Jobin Yvon) in a back scattering geometry with a radiation of He-Ne laser (λ=632.8 nm, $P_0$=1.3 mW). Temperature dependent Raman measurements were performed in back-scattering geometry with a long range 50x objective on the film samples, glued by Ag paste on a Cu sample holder, which was attached to a TiN resistive heater. A NiCr-Ni thermocouple was used to measure the temperature of the Cu sample holder.

The in situ optical ellipsometry was measured during the growth, cooling and subsequent heating of the films by using a home-made optical setup[25] of the polarizer-modulator-sample-analyser (PMSA) type. A He-Ne laser beam (λ=632.8 nm) was focused on the sample at an incident angle close to the Brewster angle of YSZ substrate, $\theta_B$~60.5°. The ellipsometric phase shift angle, Δ, and polarization rotation angle, $\Psi$, were measured by using lock-in technique at fundamental and second harmonic frequencies, $\omega$=50 kHz and 2$\omega$, respectively ($\omega$ is the modulation frequency of the light polarization). The real and imaginary part of refraction index at $\lambda$=632.8 nm was calculated using Fresnel formulas by modelling the sample as a thin film and infinitely thick YSZ substrate (see Supporting Information (SI) as ref. 26).

The local structure of TMO films was studied by Scanning Transmission Electron Microscopy (STEM) using an FEI Titan 80-300 G2 environmental transmission electron microscope (ETEM), operated at an acceleration voltage of 300 kV. TEM lamellas were prepared by Focused Ion Beam (FIB) lift-out technique using a ThermoFischer (formely FEI) Helios 4UC instrument.

## Results and discussion

In Fig. 1a) and b) we show the XRD patterns of two representative TMO/YSZ(111) films, prepared with a slightly different precursor molar ratio Tb/Mn=1.15 and 1.20, respectively. Along with the YSZ(111) substrate peaks one can see the system of reflexes from the (000l) planes of h-TMO films, evidencing an out-of-plane epitaxial growth of the h-phase. The calculated c-axis lattice parameters c=1.1152 nm and 1.1147 nm are significantly smaller than the estimated theoretical value c=1.153 nm for a strain-free h-TMO film[23]. Moreover, these values are much closer to the theoretical value for a strained h-TMO film, $c_{th}$=1.1057 nm, obtained by first principles calculations[23]. For the cubic YSZ with lattice parameter $a_0$=0.512 nm the geometrical length of the triangle side in the YSZ(111) plane is $a_{111}=a_0*\sqrt{2}$=0.724 nm. This could lead to a huge in-plane lattice misfit strain, $\varepsilon=(a_{TMO}-a_{111})/a_{111}$~-6.6 %, estimated by taking the theoretical value of in-plane lattice constant of a strained h-TMO, $a_{TMO}$=0.676 nm[23]. Such large strain is incompatible with coherent epitaxy and should lead to strain relaxation. However, the lattice film/substrate misfit could be additionally strongly reduced at the high deposition temperature ~900-1000°C by means of the reduction of the thermal h-TMO/YSZ misfit due to a very large in-plane lattice expansion coefficient of a hexagonal manganite. Indeed, extremely large values $\alpha_a$~$40*10^{-6}$ for high temperatures, T=600-1000°C, were obtained for a similar h-$YMnO_3$[27]. They exceed significantly the thermal expansion coefficient of the YSZ substrate, $\alpha_{YSZ}$~$9*10^{-6}$ (Ref. 28), and allow to reduce the film/substrate misfit at the deposition temperature, $T_{Substrate}$=900-1000°C. To preserve the high temperature strain state of the film one has to quench it down to T≤600°C because in this temperature range the lattice expansion coefficient of the h-$YMnO_3$ drops down to $\alpha_a$~$9.5*10^{-6}$ (Ref. 27) and becomes comparable with that of the YSZ substrate. Due to the lack of experimental data on h-TMO films we are basing here on a similar prototype material h-$YMnO_3$, which should, however, reflect a typical thermal behavior of a hexagonal manganite. Thus, we can conclude the presence of a large tensile stress, ε~-3 %, in the MAD grown h-TMO/YSZ films, quenched down to 600°C.

No signs of the orthorhombic phase or other impurity phases were seen in XRD for the quenched h-TMO films. The rocking curve of the (0002) peak, shown in the inset of Fig. 1a), displays a full-width-on-half-maximum, $FWHM_{TMO}$=0.033°, which is very close to that measured for the substrate, $FWHM_{YSZ}$=0.032°, for our XRD setup. Thus, the crystalline quality of our h-TMO/YSZ(111) films is determined solely by the quality of the substrate as should be in the case of a true epitaxial growth. In-plane epitaxy of a h-TMO film was verified by the full ϕ-scan around the ($11\bar{2}2$) plane shown in the inset of Fig. 1b). One can see that the ϕ-scan obeys a six-fold symmetry and contains very narrow (FWHM~0.2°) and intense peaks, evidencing the in plane epitaxy of a hexagonal film on a cubic (111)-oriented YSZ substrate with two crystallographic film domains rotated in-plane by 60° with respect to each other. In addition, the hexagonal structure was further confirmed by the room temperature polarization dependent Raman spectroscopy. In Fig. 2a) we present polarized Raman spectra of the film, normalized to the Raman spectrum of the YSZ substrate. Being only visible in the cross-polarization (yx) configuration, the strong Raman line at ~683 $cm^{-1}$ corresponds nicely to the $A_1$ mode of the hexagonal structure as shown for hexagonal manganite films[29]. Finally, the atomic force microscopy (AFM) images of h-TMO films[26] reveal a flat and smooth surface topography of the films with calculated mean-square-roughness, RMS~0.2 nm.

In Fig. 3 we present a high resolution cross-section TEM image of a h-TMO/YSZ(111) film, showing a coherent epitaxial growth with in-plane epitaxial relationships [1000]TMO//[1-10]YSZ. The atomically sharp and flat film/substrate interface contains no signs of chemical intermixing and of misfit dislocations, thus, confirming the fully coherent epitaxy strain of h-TMO/YSZ(111) films. Along the [11-20] direction one can clearly see a typical shift of Tb ions from the atomic planes, which was also observed for other bulk hexagonal manganites[30-33] and ferrites[34] and argued to be characteristic feature of the "geometric ferroelectricity". Moreover, following the "Tb shifts" along the same atomic layer one can find regions where the characteristic "up"-"up" shift sequence (left part of Fig. 3), changes to the "down"-"down" sequence. This signals the crossing of FE domain boundary and allows one to superimpose the TEM image onto the FE domain structure. The latter contains domains with electric polarization aligned perpendicular to the film plane and with a lateral size, a~5-20 nm. The observed FE domain boundary/wall is very sharp (typically 1 u.c.), directly linking the “up”- and “down”-shifted Tb ions. The domains are elongated along the z-axis and can originate from the substrate surface, but the domain size is not necessarily correlates with the film thickness as shown in Fig. SI-2b)[26]. Occasionally, some of large domains observed in the film originate from the atomic steps at the substrate surface (see Fig. SI-2a)[26]). It is worth to mention that according to our TEM study the very first 3-4 u.c. of h-TMO do not reveal “Tb-shifts”, indicating the absence of ferroelectricity in very thin layers close to the substrate. The reason could be a dimensional effect, i.e. a thermodynamically stable FE domain needs a space which cannot be arbitrary small, limiting the smallest size to 3-4 u.c. Our results on strain-stabilized TMO films agree well with the observation of vanishing of the improper ferroelectricity in 3 u.c. thick $YMnO_3$ films[35].

In Fig. 2b) we present the Raman spectra of the h-TMO film measured at fixed temperatures (the spectra are shifted for clarity) taken by heating and cooling (not shown) the sample in the temperature range, T=293-973 K. The Raman spectra in Fig. 2b) of the film sample were first corrected by the Bose factor, $n+1=\frac{1}{1-e^{\hbar\omega/kT}}$, and then normalized to the strongest substrate peak at ~613 $cm^{-1}$ as can be seen also in Fig. 2a). Qualitatively, one can see in Fig. 3b) a strong temperature dependent Raman behavior with suppression of the film $A_1$ peak, started already from room temperature, and complete vanishing of the film peak at T>873 K. The subtracted (sample-substrate) Raman spectra presented in Fig. 4a) allowed us to analyze quantitatively the temperature behavior of a single $A_1$ film peak by fitting the difference spectra at all measured temperatures by Lorentz lines as shown in Fig. SI-3[26]. One can see in Fig. 4b) that the intensity, quantified by the area under the $A_1$ peak and its height, decrease steeply for T=300-450 K and then continuously goes down to zero for T=500-773 K, indicating the existence of a structural phase transition at about 800 K. A similar temperature behavior with vanishing of the $A_1$ peak for $T>T_C$=1020 K was reported in h-$LuFeO_3$ film[16]. The position of the $A_1$ peak in our strain-stabilized h-TMO film shifts almost linearly from ~683 $cm^{-1}$ at room temperature down to ~668 $cm^{-1}$ at 773 K (see the inset in Fig. 4b)). This linear temperature behavior agrees with that measured in h-$YMnO_3$ bulk samples[15]. Note, that the $A_1$ peak in bulk samples of h-$YMnO_3$[15] did not vanish for $T>T_C$~1000 K, but rather changes the symmetry to the $A_{1g}$. However, for T>800 K we cannot further resolve the Raman peak from our 50 nm thick h-TMO film because of very small peak intensity, which in addition superimposes with the Raman contribution of the substrate. After cooling down the film to room temperature the intensity and position of the $A_1$ Raman peak retain to the same level as before heating, indicating no sizable strain relaxation effects in this film. Thus, the temperature dependent Raman study indicates that a structural phase transition into a paraelectric $P6_3/mmc$ structure occurs in the h-TMO/YSZ(111) film around $T_C$~800 K.

To get an additional insight into the structural phase transition in hexagonal h-TMO films we studied the temperature dependent optical ellipsometry for T=300-1200 K. In Fig. 5a) the time dependences of the ellipsometric phase shift, $\Delta(t)$, and polarization rotation, $\Psi(t)$, angles are shown by cooling the film. Using double-side polished YSZ substrate we were able to see the oscillations in $\Delta(t)$ and $\Psi(t)$ due to the optical interference of the laser beam reflected from the top and bottom surface of the plane parallel substrate, the thickness of which is changing due to thermal expansion. This allows us to measure the temperature of the sample by calculating the number of oscillations and calibrating the temperature by using a pyrometer, namely, T=1200 K was achieved at the highest current I=28 A. One can see in Fig. 5 that both $\Delta(t)$ and $\Psi(t)$ show the following characteristic features: a) continuous and pronounced changes of about 3-5° at the beginning and at the end of the time scan and 2) a steep step-like change of the signal, especially pronounced in $\Delta(t)$, changing by 10°, in the middle part of the scan. In Fig. 5b) the time scale was recalculated into the temperature scale by using the relation T(t)=293 K+18 K*t, where the initial point of the x-scale at t=0 corresponds to the finally achieved room temperature T=293 K (20°C). Apparently, both ellipsometric angles $\Delta(T)$ and $\Psi(T)$ change drastically for T=600-800 K, indicating the phase transition takes place within the same temperature interval as for the Raman spectroscopy (see Fig. 2 and 4). In addition, the evaluated temperature dependence of the real part of refractive index, N(T), is shown in Fig. 5b), obtained by assuming a simple optical model[26] of a homogeneous film grown on YSZ substrate having $N_{YSZ}$=2.15 at $\lambda$=632.8 nm (see Ref. 36). The refractive index for a h-TMO film, n~2.12, at room temperature, being comparable to that obtained for $YMnO_3$ (Ref. 37), decreases drastically from n=2.12 to n=2.0, i.e. by 6 %, when the film undergoes the FE/PE phase transition. The endpoint of this transition in N(T) can be assigned to the Curie temperature, $T_C$=800 K, which agrees nicely with the vanishing of the $A_1$ Raman peak in Fig. 4b).

With thermal expansion coefficient ~$10^{-5}$ and temperature change ~$10^3$ K, the estimated change of the film thickness d=50 nm at room temperature by heating/cooling is 0.5 nm, i.e. ~ unit cell. Such small changes cannot result in the observed thermal behavior (see Fig. 5). The origin of the observed ellipsometric behaviour should be related to the optical properties and electronic structure of h-TMO, which apparently change at the structural phase transition. The photons from He-Ne laser with energy E=1.95 eV corresponds to a strong and narrow absorption band, the centre of which lies at E=1.75 eV as shown in the inset of Fig. 5 for h-$TbMnO_3$ (see Ref. 38). According to the theory[39], the origin of this absorption band in hexagonal manganites is the inter-band charge transfer excitation from the occupied oxygen O2p states into empty Mn3d states. Hence, we can conclude that the p-d charge transfer demonstrates a distinct change at the structural phase transition in h-TMO film. Note, that the present study deals with a monochromatic excitation only and, thus, we cannot directly measure the changes of spectral weight transfer at the phase transition. In addition, the blue shift of the absorption peak by decreasing temperature is known only for low temperatures, T=300-15 K[38]. Here we can speculate, that, a modification of the electronic band structure during the $P6_3/mmc$ (PE) to $P6_3cm$ (FE) structural phase transition influences a charge transfer from the O2p to Mn3d states. Note, that charge transfer from the Y-$O_T$ (apical oxygen) bonds to the $O_T$-Mn bonds of Mn3*d*-O2*p* hybridized states, was found to be responsible for the enhanced ferroelectricity during the annealing-induced orthorhombic-hexagonal phase transition in bulk $YMnO_3$ (Ref. 40).

In the bottom panel of Fig. 3 we present the in situ TEM heating experiments, performed to monitor the changes of atomic structure and morphology of ferroelectric domains in the h-TMO with increasing temperature. Electron diffraction patterns, collected at every 5 K by heating, do not show any changes for T=300-400 K. However, already at T=405 K new superstructure spots, marked with the white arrow in the inset in Fig. 3c), show up. Finally, at T=410 K all spots originating from the ferroelectric hexagonal $P6_3cm$ structure along with Tb shifts disappear and another hexagonal $P6_3mcm$ structure is formed similar to the structural transformation described for the bulk h-$YMnO_3$[41]. One has to note that this structural transformation is irreversible in the lamella of initially h-TMO film and, is most probably, related to the temperature-induced full stress relaxation in a very thin (~20 nm) TEM lamella with very large aspect ratio 1x10. This results in the formation of a paralelectric phase with a hexagonal structure in the TEM lamella already at 410 K<<$T_C$~800 K. Considering a strong decrease of the intensity of $A_1$ Raman peak for T=300-450 K (see Fig. 4b)), one can suggest that a partial strain relaxation by heating can also take place in the whole film sample. However, in the whole film sample such strain relaxation seems to be reversible.

The result of irreversible strain relaxation due to the "Raman annealing" of the h-TMO film is shown in Fig. SI-4a)[26]. One can see that the c-lattice parameter changes after T-dependent Raman measurements from c=1.115 nm (as prepared) to c=1.109 nm (after Raman annealing). Note, that the difference $\Delta c$=0.006 nm is very close to the difference between the c-axis of the FE ($P6_3cm$) and PE ($P6_3/mmc$) structures of $YMnO_3$[42]. Interestingly, by additional heating and quenching the film from the deposition temperature down to ~600°C the as prepared state with c=1.116 nm can be recovered (see Fig. SI-4a)[26]). This is also supported by the recovering of the Raman intensity at room temperature (not shown) after quenching. Remarkably, the strain relaxation state in the relatively thin h-TMO films with thickness d~50 nm can be also reversible as the film after heating and cooling during Raman measurements (2-3 hours) comes finally back to the original Raman spectrum (see Fig. 4b)). We believe that the difference between the reversible (R) and irreversible (IR) thermal behaviour of h-TMO films could be related to the deviations in their stoichiometry as the R- and IR-films were prepared by using slightly different precursor molar ratio, Tb/Mn=1.15 and 1.2 ratios, respectively.

To correlate the global XRD structural behavior (Fig. SI-4[26]) with atomic scale changes of the h-TMO structure induced by heating, the cross section TEM specimens were cut from the R- and IR-films after "Raman annealing". The HRSTEM HAADF images (Fig. SI-5a)-d)[26]) were collected using identical parameters of the electron microscope for both films. The original images were FFT filtered using selected space frequencies and their contrast was enhanced (see Fig. SI-5[26] insets in a) and c)). The filtered image of the IR-film (Fig. SI-4d) demonstrates an inhomogeneous microstructure with FE domains, intermixed by PE domains (flat Tb planes) with ~3-7 nm size. In contrast the R-film displays a homogenous microstructure with FE domains with a lateral size of more than 30 nm (Fig. SI-5 b)). Moreover, the interface layer with PE structure is by ~50 % thicker in the IR-film (Fig. SI-5c)) as compared to that in the R-film (Fig. SI-5a)). The inhomogeneous local structure of the IR-film with coexisting FE and PE domains, induced by the irreversible temperature transformation, qualitatively agrees with changes in XRD and Raman data. Note that the c-lattice constant of the R-film with homogeneous microstructure (see Fig. SI-5b)[26]) did not change after the Raman annealing (see Fig. SI-4b)[26]). At the same time, considering the fact that the intensity of the $A_1$ peak strongly decreases already for $T \ll T_C$ and the temperature dependence of the $A_1$ peak (see Fig. 4b) looks very different from that expected for a temperature-driven phase transition[42], one can suggest that the strain rather than temperature governs the structural behaviour of h-TMO film for low temperatures far away from $T_C$. This points out onto a high sensitivity of the ferroelectric domain structure to the changes of the epitaxy strain, originated from the YSZ(111) substrate.

In conclusion, high quality epitaxial thin films of hexagonal $TbMnO_3$ were grown on YSZ(111) substrates by using MAD technique. Compared to the previously reported PLD-grown films they display a perfect crystallinity and atomically sharp film/substrate interfaces. Ferroelectric domains, originated from the shift of Tb ions from the atomic plane, with sharp domain boundaries were visualized by means of HRTEM. Structural phase transition at $T_C$~800 K was determined by the temperature dependent Raman spectroscopy and optical ellipsometry; the latter indicates the change of charge transfer at the structural phase transition. Ferroelectric/structural domains in epitaxially stabilized h-TMO films depend drastically on the stress macroscopic state of the film.

**Acknowledgements**

R.M. acknowledges financial support from Erasmus Plus programme, European Union, Georg-August-Universität Göttingen and IISER Pune. V.R., U.R. S.M. and V.M. acknowledge financial support from Deutsche Forschungsgemeinschaft via SFB 1073 (TP A02, Z02). Authors thank M. D. Bongers for XRD phi-scan measurements. The use of equipment in the "Collaborative Laboratory and User Facility for Electron Microscopy" (CLUE) www.clue.physik.uni-goettingen.de is gratefully acknowledged.

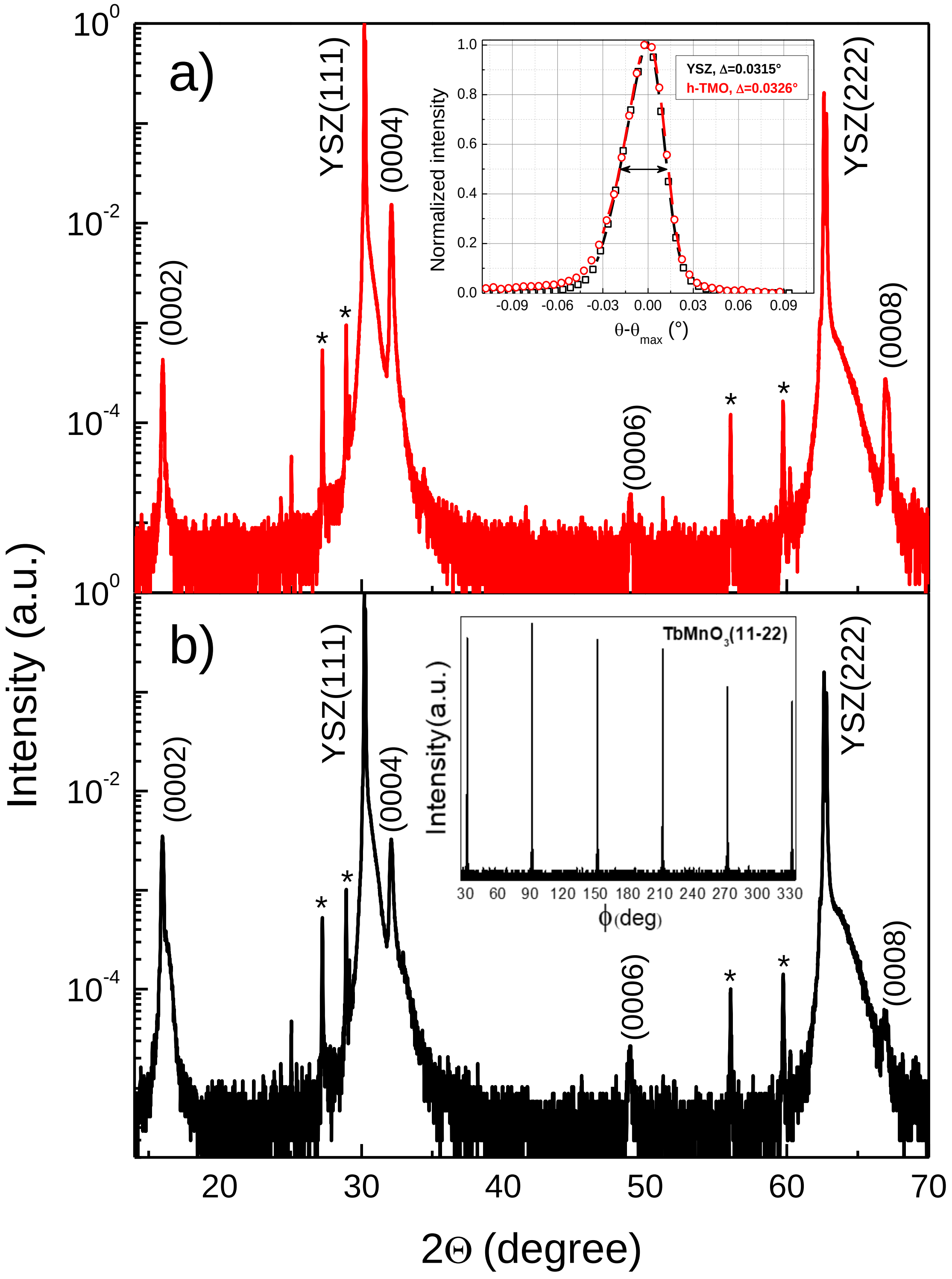

Fig. 1: XRD Θ-2Θ scans for two representative h-TMO/YSZ(111) films, showing the (000l) out-of-plane epitaxy, with calculated out-of-plane lattice parameters, c=1.1152 nm and 1.1147 nm. The insets demonstrate rocking curves of h-TMO(0002) and YSZ(111) peaks (a) and a phi-scan of the film (11-22) peak (b), both evidencing a perfect in-plane epitaxy with hexagonal symmetry.

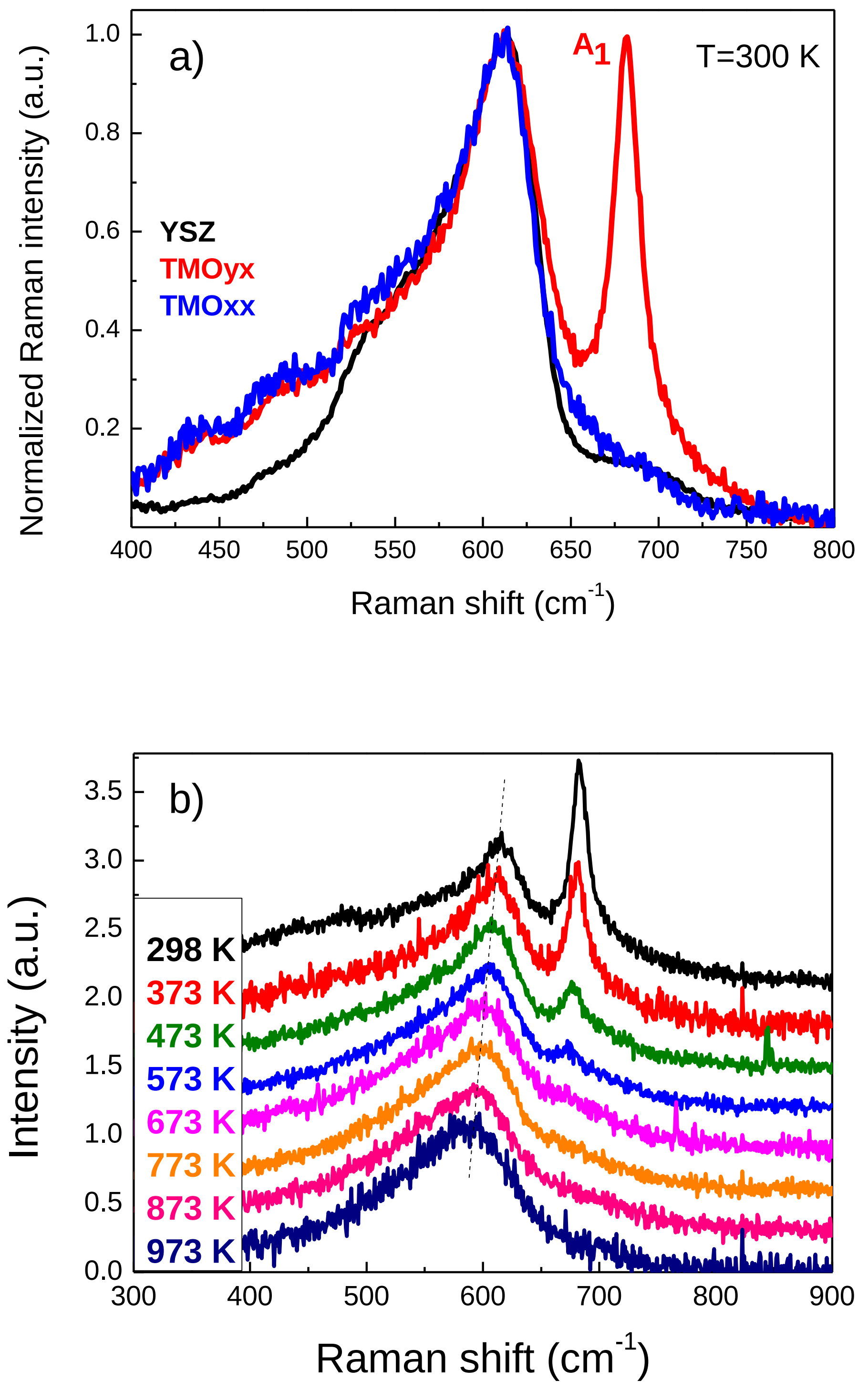

Fig. 2: Raman spectra of a TMO/YSZ(111) film measured at room temperature in XX (red) and XY (blue) polarization geometries confirm the hexagonal structure of TMO films; b) Temperature dependent unpolarised Raman spectra, vertically shifted for clarity. The spectra of the films are normalized to the Raman spectrum (black) of a blank YSZ(111) substrate with maximum at ~613 $cm^{-1}$ at room temperature.

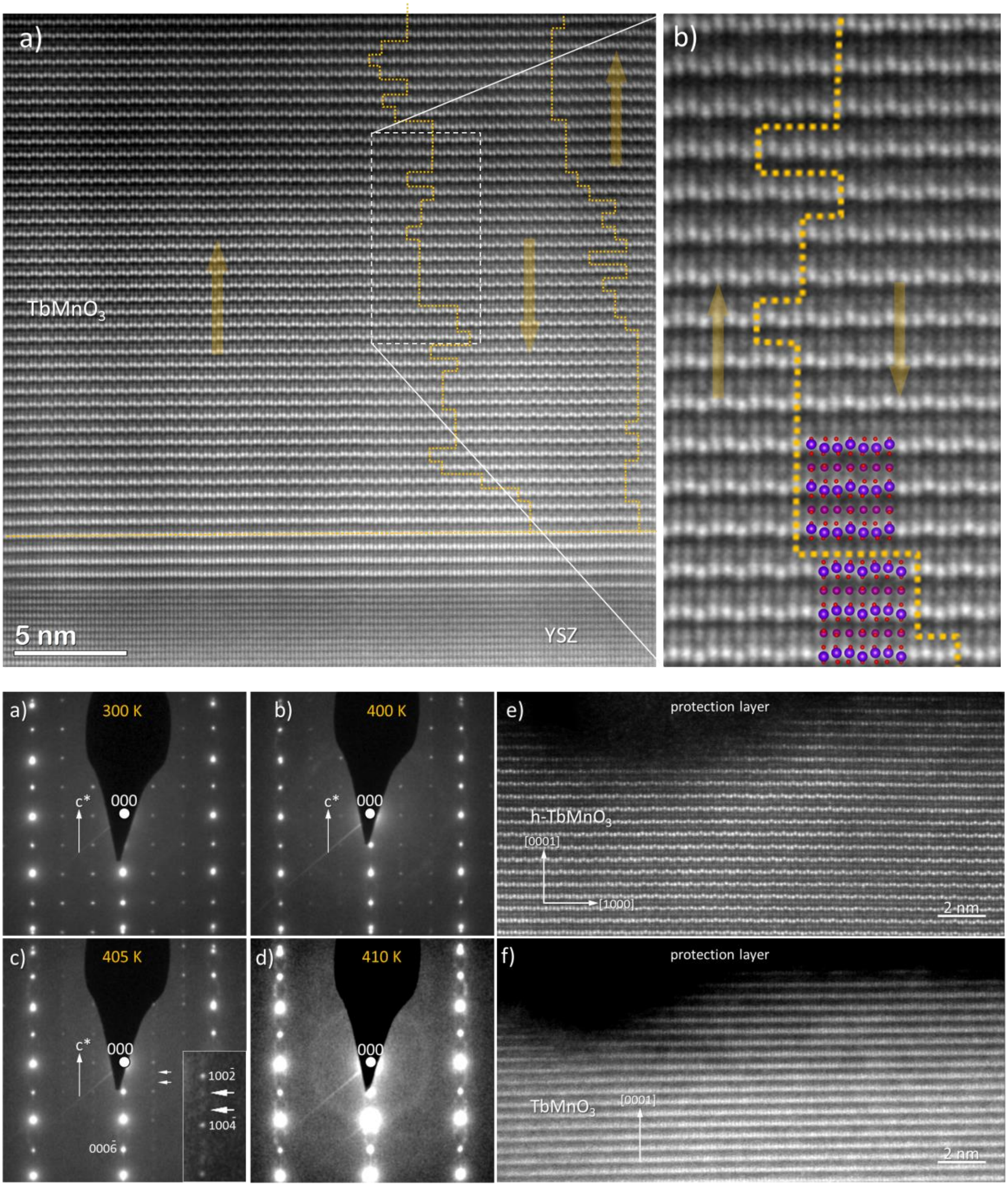

Fig. 3. Top panel: a) High resolution cross-section TEM images of a h-TMO/YSZ(111) film measured along the [11-20] crystallographic direction of the film with characteristic buckling (rumpling) of Tb atoms. One can also see an atomically smooth and flat film/substrate interface as well as the absence of Tb-rumpling in the very first 4 u.c. of the film. b) In the zoomed part of Fig. 3a) one can clearly see the characteristic regions with the "up"- (left part) and "down"-rumpling (right part) of Tb ions, separating FE domains with a typical size varying in the range 5-20 nm. Bottom panel: (a)-(d) Electron diffraction and high resolution STEM images, collected at different temperatures upon heating of the TEM lamella up to 410 K. The initial hexagonal ($P6_3cm$) structure with Tb-shift (e) was irreversibly transformed into a hexagonal ($P6_3mcm$) structure with planar Tb-planes (f).

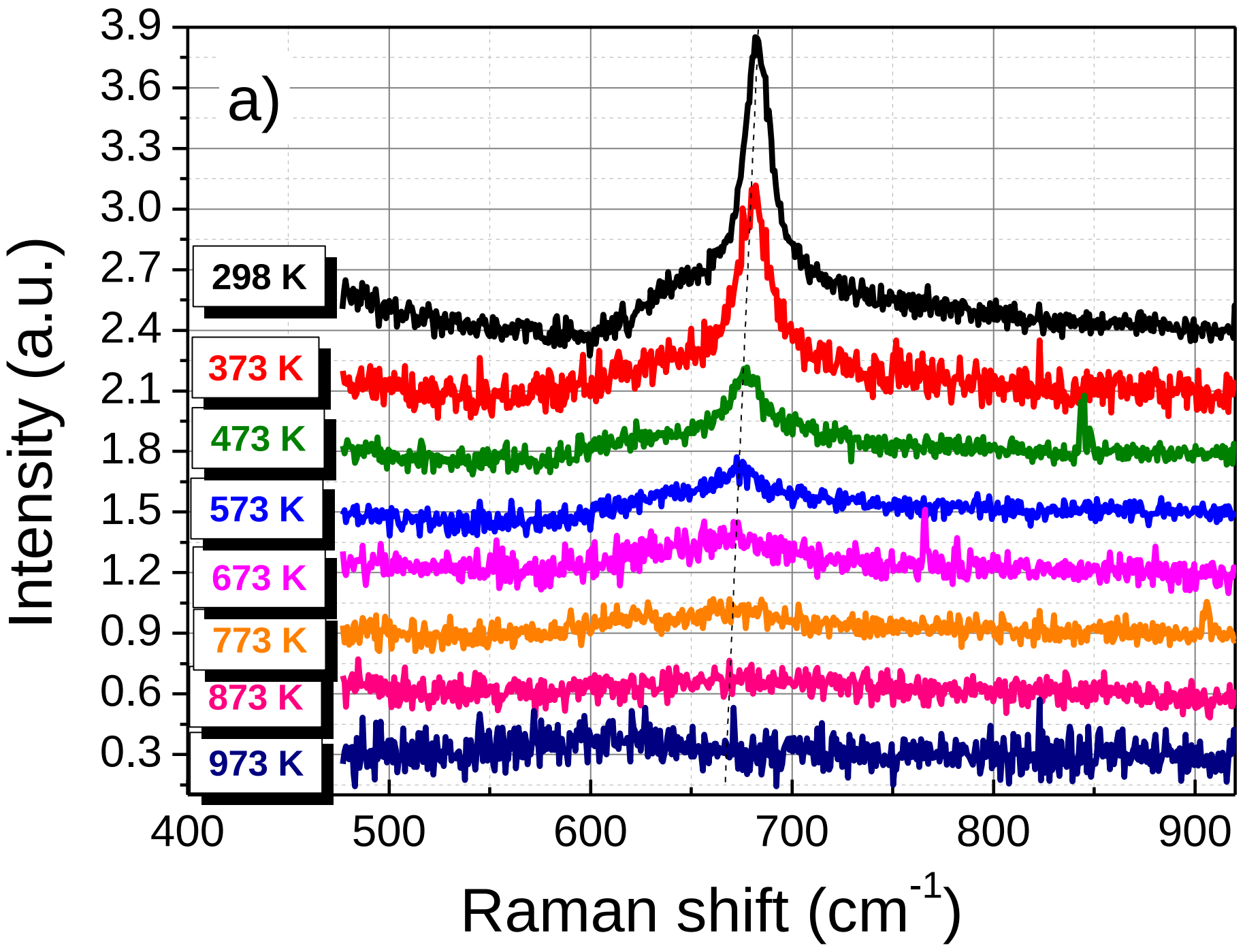

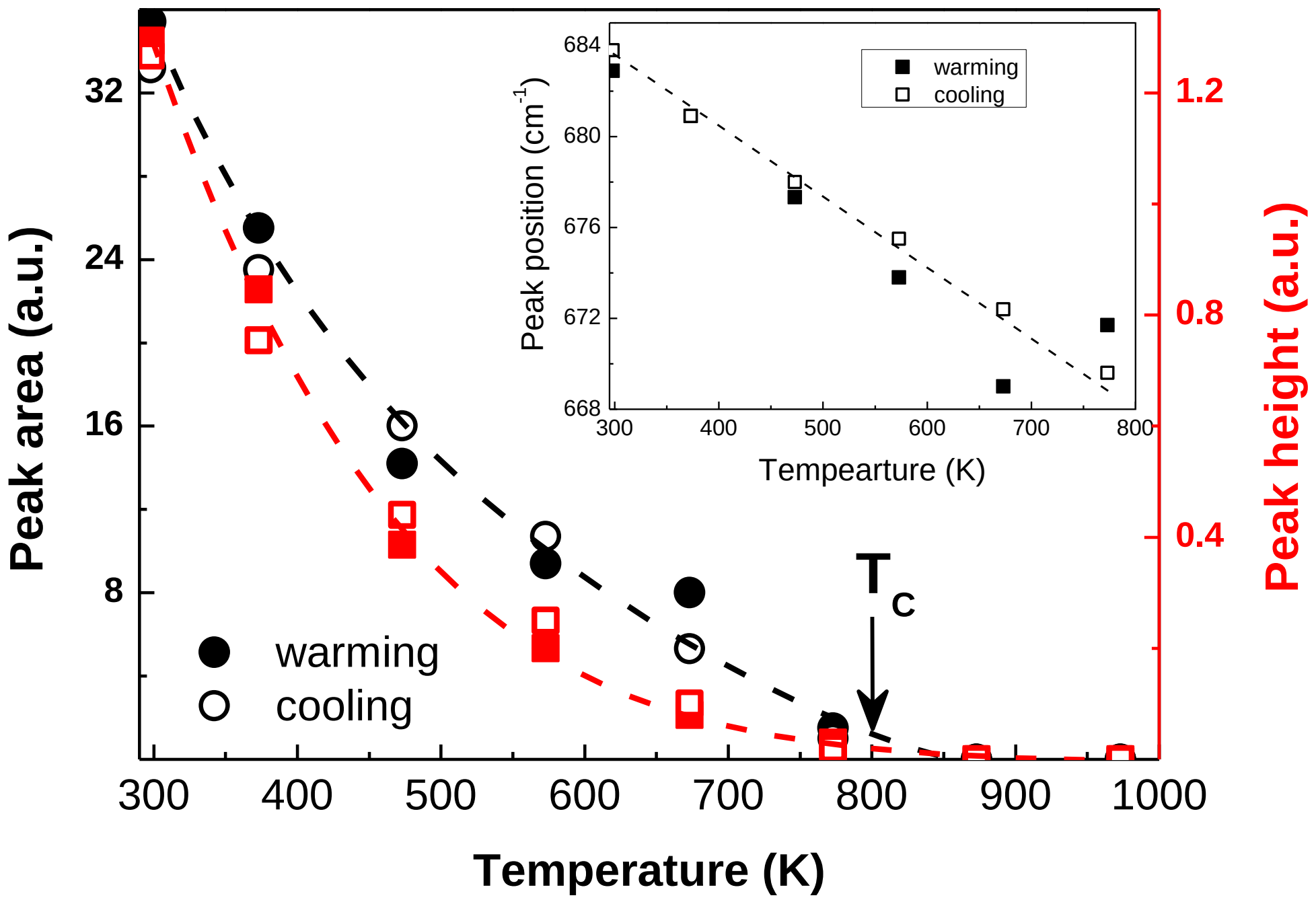

Fig. 4: a) The evaluated temperature dependent Raman spectra of the h-TMO film; b) The intensity (area under the peak, left scale) and the height of the $A_1$ peak (right scale) by warming (closed symbols) and cooling (open symbols) indicate the reversible temperature behaviour of the $A_1$ peak. A phase transition about 800 K is indicated by the vanishing of the $A_1$ peak. The peak position as a function of temperature is shown in the inset in b).

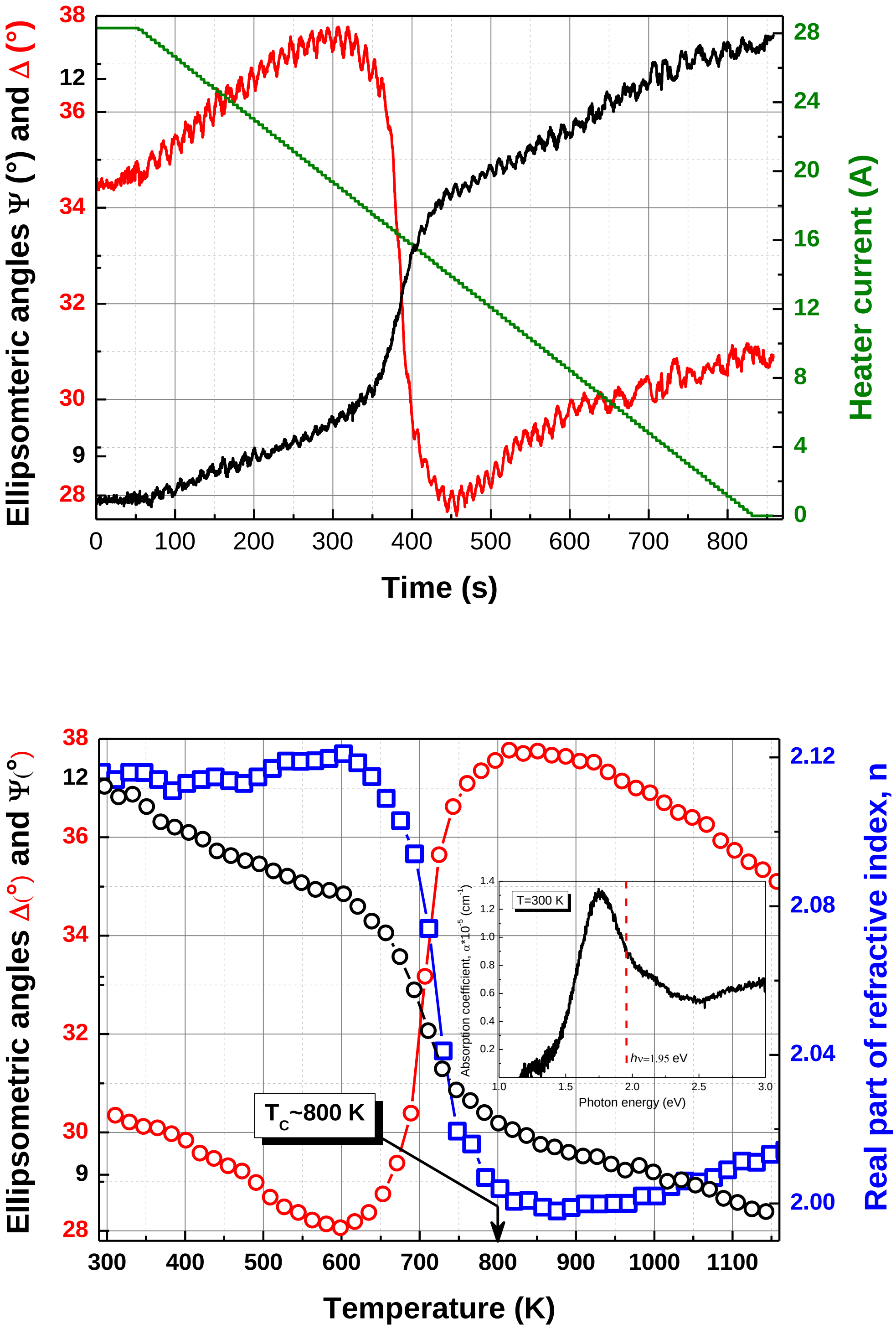

Fig. 5 Time (a) and temperature (b) dependences of ellipsometric phase shift, Δ, and polarization rotation, Ψ, angles as well as of the calculated refractive index, n. A distinct step-like decrease of N by 6 % indicates the phase transition, the end point which marks $T_C$~800 K.

# Supporting Information

## Strain-Driven Structure-Ferroelectricity Relationship in hexagonal $TbMnO_3$ Films

R. Mandal[1,2], M. Hirsbrunner[1], V. Roddatis[3], L. Schüler[1], U. Roß[3], S. Merten[1] and V. Moshnyaga[1*]

[1]*Erstes Physikalisches Institut, Georg-August-Universität Göttingen, Friedrich-Hund-Platz 1, 37077 Göttingen, Germany*

[2]*Department of Physics, Indian Institute of Science Education and Research, Pune 411008, India*

[3]*Institut für Materialphysik, Georg-August-Universität Göttingen, Friedrich-Hund-Platz 1, 37077 Göttingen, Germany*

**E-mail:* vmosnea@gwdg.de

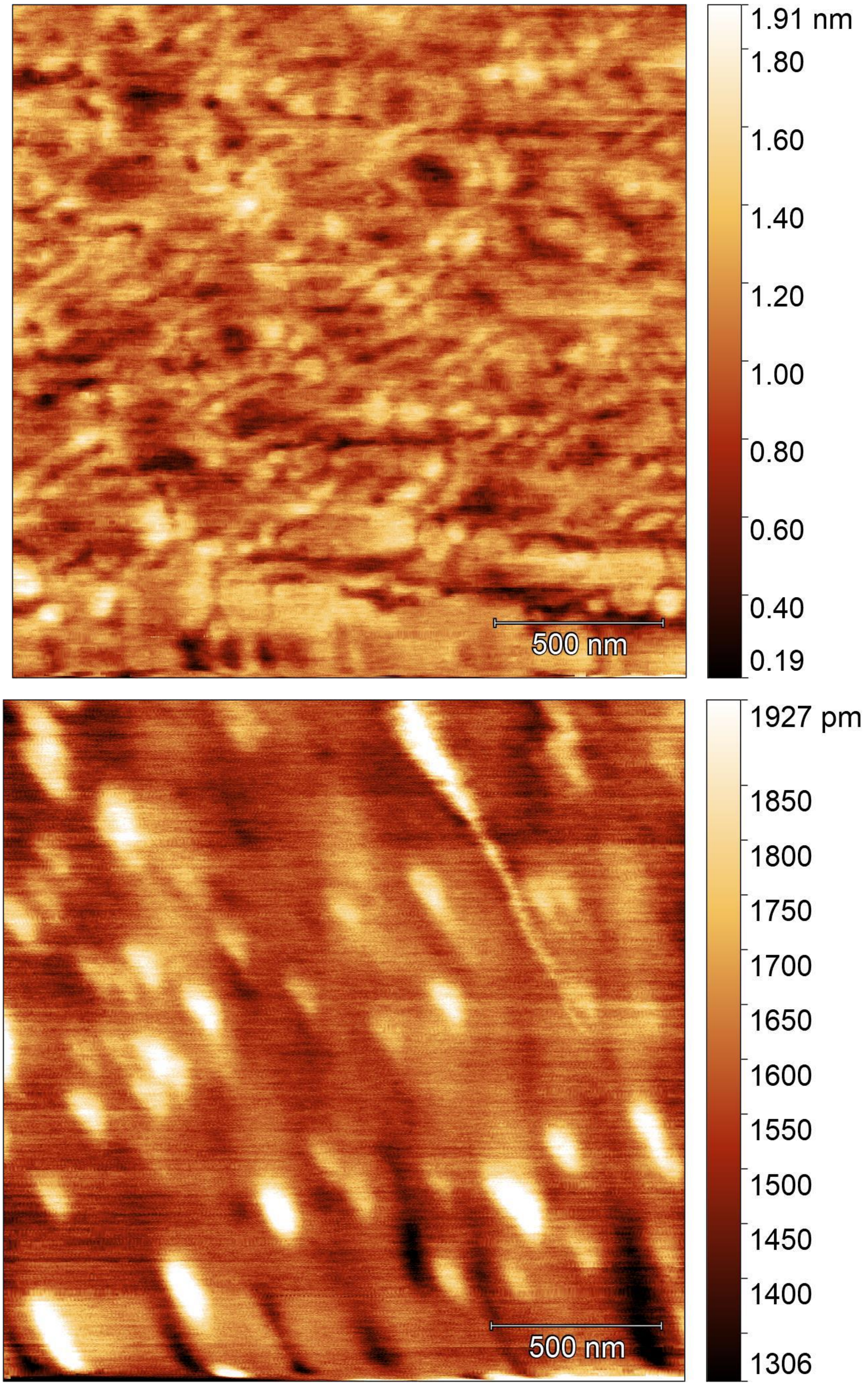

Fig. SI-1 AFM images of the two representative h-TMO films with Tb/Mn precursor molar ratio 1.1 (a) and 1.15 (b). One can see a very smooth and flat surface with evaluated RMS values ~0.3 (a) and 0.4 (b) nm.

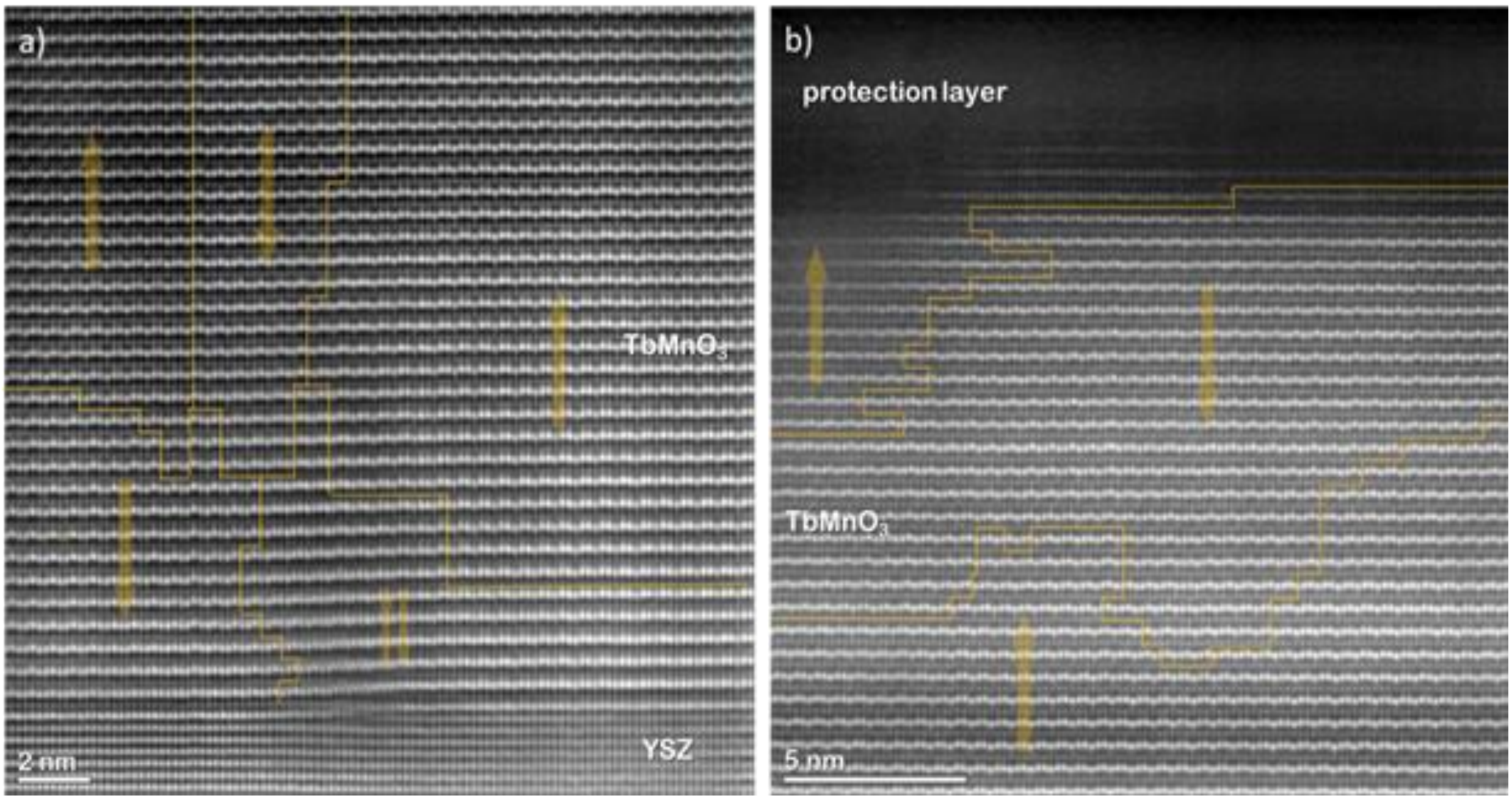

Fig. SI-2 (a) HRSTEM image of $TbMnO_3$/YSZ interface showing the origin of FE domain at the atomic height step. (b) HRSTEM image of the top part of the $TbMnO_3$ film. A complicated domain structure is shown by dotted lines marking FE domain walls and pointing “up” and “down” polarization vectors.

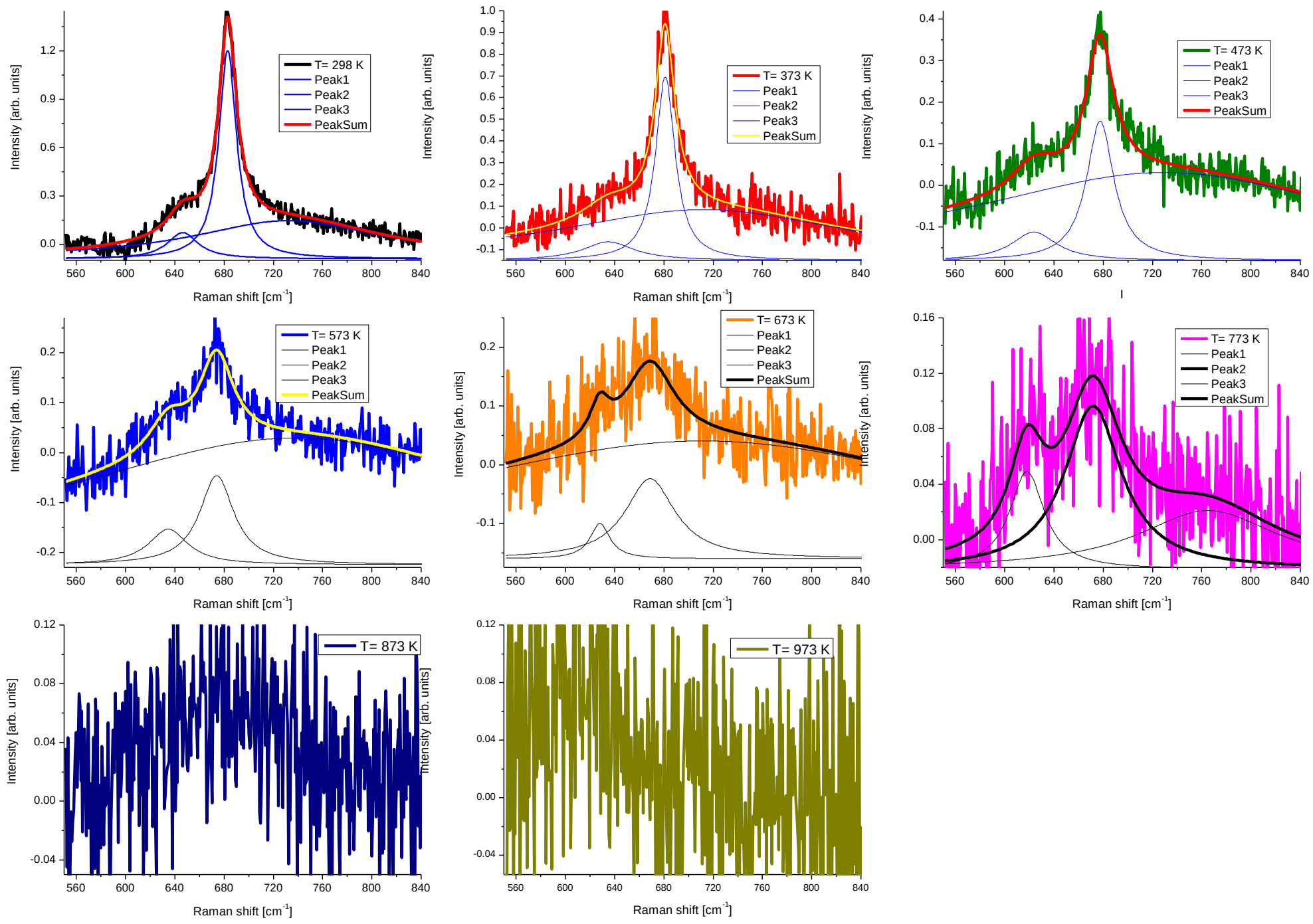

Fig. SI-3 Evaluated Raman spectra of the film at different temperatures. All spectra for T=300-773 K can be fitted by three Lorentz peaks, originated from the YSZ substrate (peak 1 ~613 $cm^{-1}$ at room temperature (RT) and peak 3 due to not ideal substrate subtraction) and $A_1$ peak from the film (peak 2 @683 $cm^{-1}$ by RT).

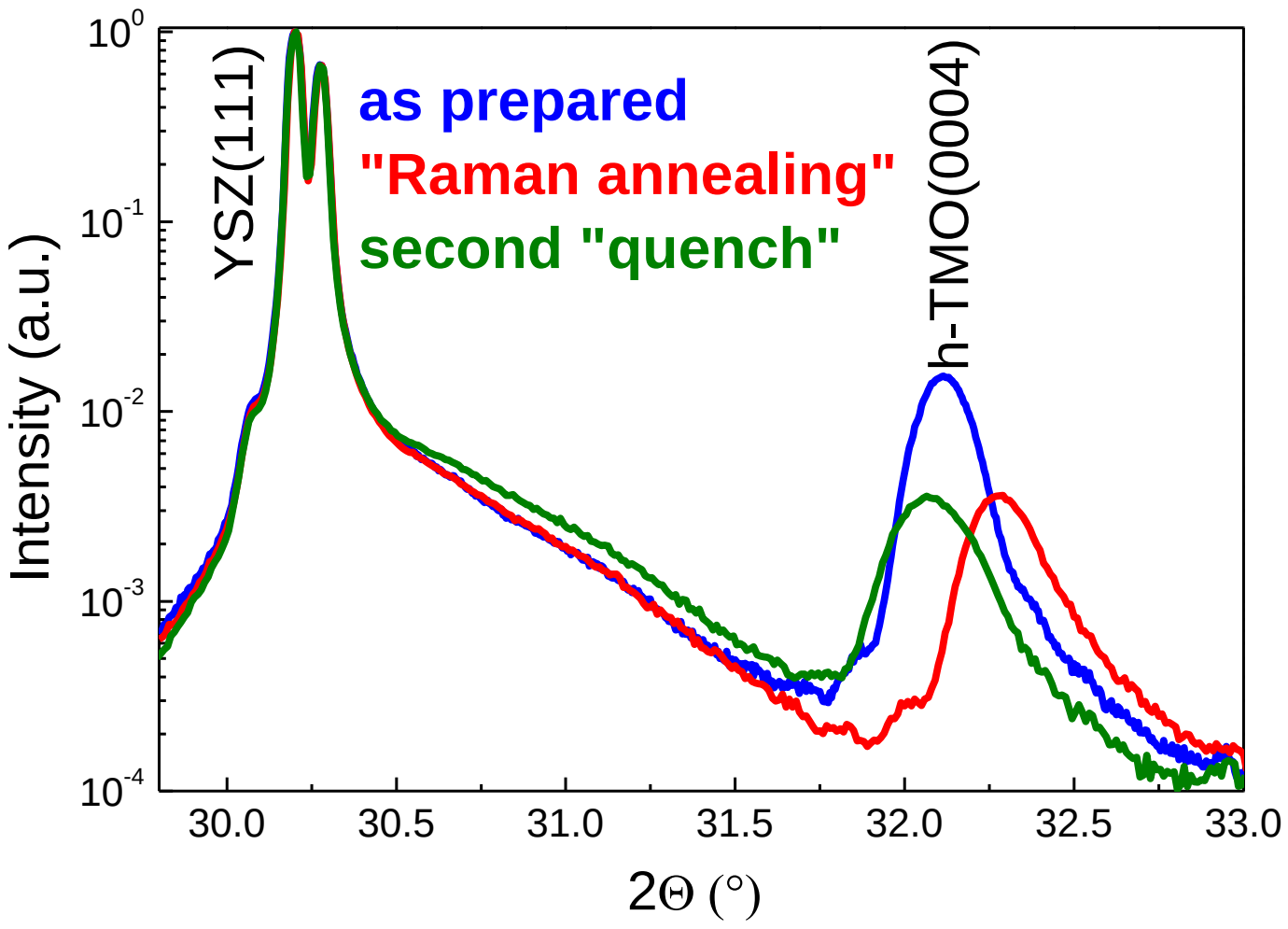

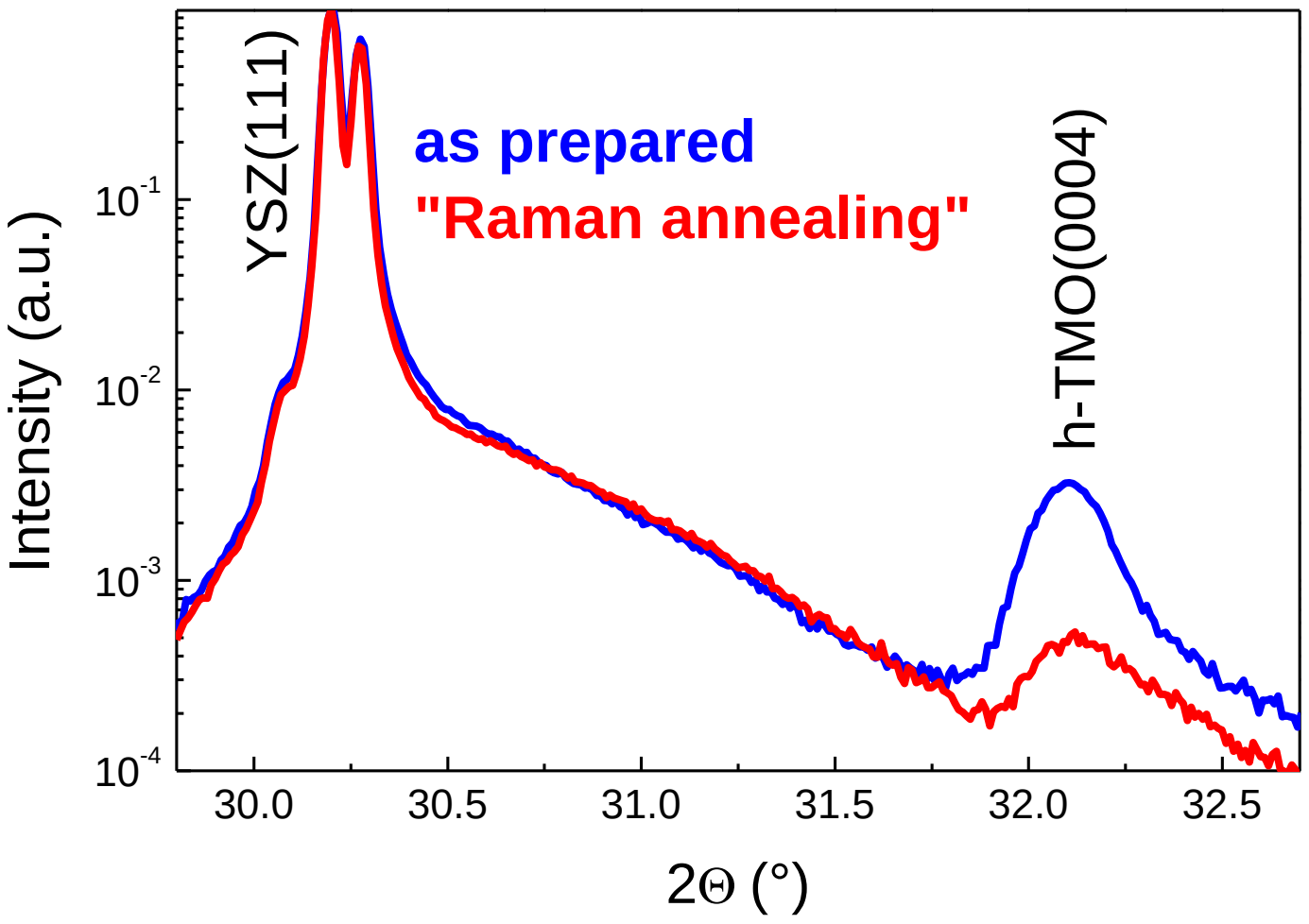

Fig. SI-4 XRD patterns in the region of the h-TMO(0002) peak of h-TMO films with a) irreversible thermal behavior with decrease of the c-axis lattice parameter after annealing during Raman T-scans for T=293 K-973 K-293 K (red curve) and b) with a reversible thermal behavior, indicated by do not changing c-lattice parameter. One can "reanimate" the h-TMO film by an additional heating up to the deposition temperature, $T_{sub}$~1000 °C, and subsequent quenching it down to 600°C in 1 min (green curve).

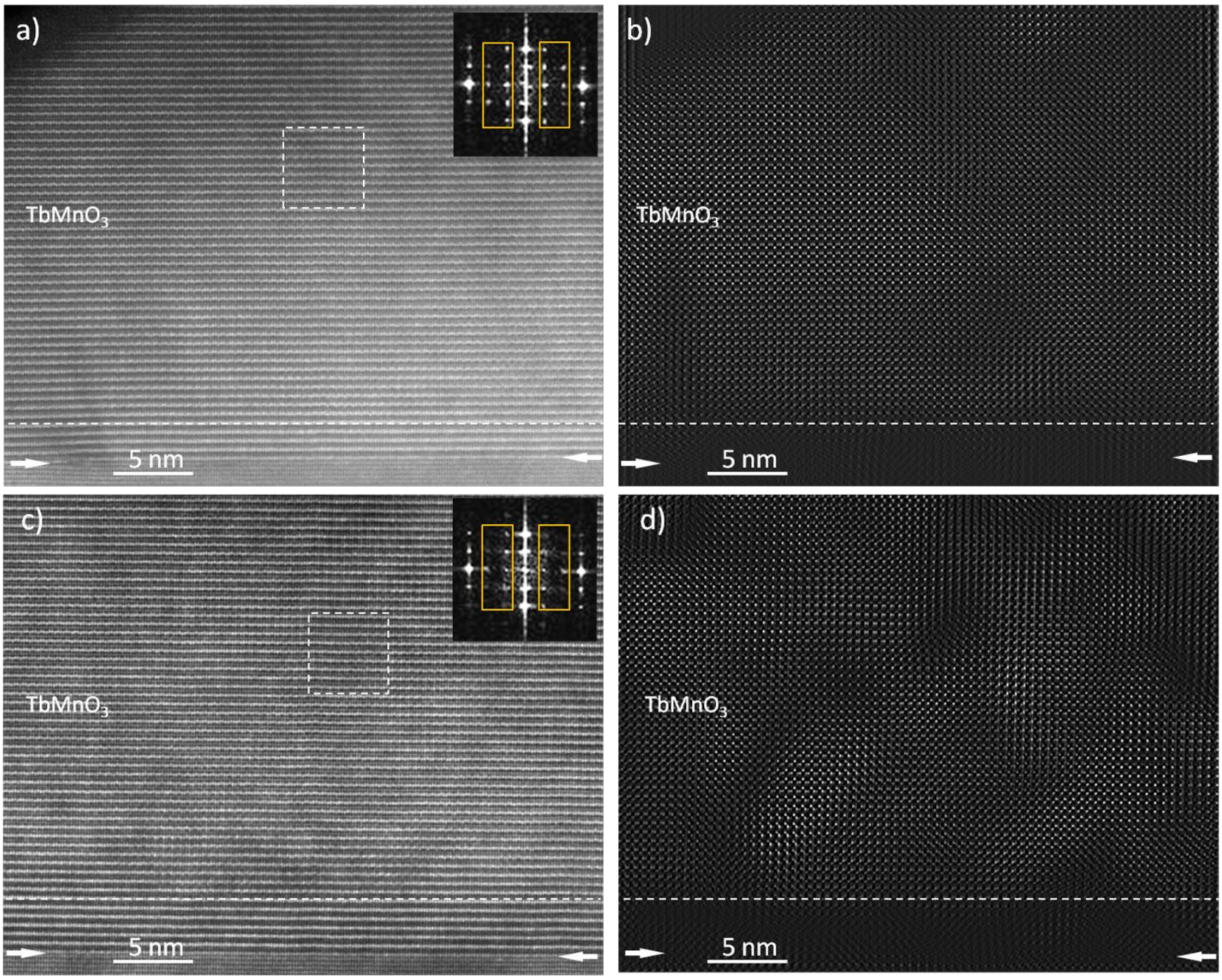

Fig. SI-5. Original and filtered HRSTEM images of the two $TbMnO_3$ films, showing the reversible (a, b) and irreversible (c, d) behavior after temperature-dependent Raman measurements. The insets show two FFT patterns from the selected areas marked with the white line. The spots inside the selected areas in the FFTs (orange contours) were used for the filtering to reveal the distribution of PE and FE phases. The FE phase demonstrates sharper and brighter spots because of pronounced up and down shifts of Tb ions, while the PE phase demonstrates almost flat Tb planes.

**Calculation of the refractive index from the measurements of optical ellipsometry**

The ellipsometry measurements yield the ratio of the complex reflection coefficients $r_p$ and $r_s$ for light polarized parallel and perpendicular to the normal vector of the sample surface. This ratio is often represented by the angles $\Psi$ and $\Delta$:

$$\rho = \frac{r_p}{r_s} = \tan\Psi \exp(i\Delta) \tag{1}$$

The sample is situated in ambient air ($N_0 = 1$) and modelled as a thin film with thickness *d* and complex refractive index $N_1 = n - ik$ on an infinitely thick substrate with refractive index $N_2 = 2.15$ [1]. The reflection coefficients for the light with wavelength $\lambda$ and angle of incidence $\theta_0$ are given by the equation:

$$r_{p,s} = \frac{r_{01p,s} + r_{12p,s}\exp(-2i\beta)}{1 + r_{01p,s}r_{12p,s}\exp(-2i\beta)} \tag{2}$$

where $\beta = \frac{2\pi d}{\lambda} N_1 \cos\theta_1$ and $r_{01p,s}$ and $r_{12p,s}$ are the Fresnel coefficients for the reflection at the air/film and film/substrate interface [2]:

$$r_{jkp} = \frac{N_k\cos\theta_j - N_j\cos\theta_k}{N_k\cos\theta_j + N_j\cos\theta_k}$$

$$r_{jks} = \frac{N_j\cos\theta_j - N_k\cos\theta_k}{N_j\cos\theta_j + N_k\cos\theta_k}$$

Using Snell's law $N_0\sin\theta_0 = N_1\sin\theta_1 = N_2\sin\theta_2$ the angles of the light beam relative to the surface normal can be calculated in each layer.

With the known wavelength $\lambda = 632.8$ nm, incident angle $\theta_0 = 60.5°$ and film thickness d=50 nm only the real part *n* and imaginary part *k* of the refractive index of the thin film remain unknown. To obtain those values from the measured ellipsometric angles $\Psi_m$and $\Delta_m$ the system of equations

$$\Psi(n,k) - \Psi_m = 0$$
$$\Delta(n,k) - \Delta_m = 0$$

is solved by the method of nonlinear least squares where $\Psi(n,k)$ and $\Delta(n,k)$ are calculated using equations (1) and (2).